# Ultra-wide plasmonic tuning of semiconductor metasurface resonators on epsilon near zero media

Prasad.P.Iyer[1], Mihir Pendharkar[1], Chris J. Palmstrøm[1,2], Jon A. Schuller [1]

1. Electrical and Computer Engineering Department, University of California Santa Barbara, CA
2. Material Science and Engineering Department, University of California Santa Barbara, CA

**Fully reconfigurable metasurfaces would enable new classes of optical devices that provide unprecedented control of electromagnetic beamforms. The principal challenge for achieving reconfigurability is the need to generate large tunability of subwavelength, low-Q metasurface resonators. Here, we demonstrate large refractive index tuning can be efficiently facilitated at mid-infrared wavelengths using novel temperature-dependent control over free-carrier refraction. In doped InSb we demonstrate nearly two-fold increase in the electron effective mass leading to a positive refractive index shift (Δn>1.5) far greater than conventional thermo-optic effects. In undoped films we demonstrate more than 10-fold change in the thermal free-carrier concentration producing a near-unity negative refractive index shift. Exploiting both effects within a single resonator system—intrinsic InSb wires on a heavily doped (epsilon near zero) InSb substrate—we demonstrate dynamically tunable Mie resonances. The observed larger than line-width resonance shifts (Δλ>1.5µm) suggest new avenues for highly tunable and reconfigurable mid-infrared semiconductor metasurfaces.**

Metasurfaces exploit arrays of subwavelength resonators with highly confined and enhanced electric fields to control the phase, amplitude and polarization of light[1,2]. These resonators typically comprise nano-antennas that enhance the electric fields based on 1) plasmonic (metallic) resonances[3] or 2) high-index low-loss Mie[4,5] (dielectric) resonances. Static, geometry-based control of resonator size, shape and material has enabled a variety of applications including miniaturized optical components[1,6], enhancement of non-linear properties[7,8] etc. Reconfigurable control over resonator properties is an emerging challenge in the field of metasurfaces[9,10]. Fully reconfigurable metasurfaces require shifting resonances by at least one line-width[11] to extract sufficiently large phase shifts[12,13] for e.g. beam steering or focusing. Tunable metasurfaces based on refractive index tuning of graphene[14,15], phase change materials[16,17], free carrier refraction[13,18], and MEMS[19] have been demonstrated, but lack the tuning range needed to achieve fully reconfigurable and continuous 2π phase shifts.

Semiconductor Mie resonators can have negligible absorption, large radiative coupling[5,20], and large tunability, making them ideal candidates for reconfigurable metasurfaces. Since typical resonators have low Qs (~3-6)[4,21], they require a large modulation of the refractive index (Δn ≥ 1)[22]. At mid-infrared wavelengths, free carrier modulation in semiconductors provides a viable route to achieve such large refractive index changes.[18] Traditionally this is achieved by modulating the plasma frequency ($\omega_p$) by optically[22–24] or electrically[12,25] pumping the metasurface to change the carrier concentration ($N_{free}$) for a fixed electron effective mass ($m_e^*$). Here we demonstrate two distinct approaches for achieving thermally tunable, larger-than-unity refractive index shifts in InSb metasurface resonators: 1) tuning the free carrier density ($N_{free}$) and 2) tuning the electron effective mass ($m_e^*$).

InSb forms an ideal thermally tunable system due to its low bandgap ($E_g = 0.17 eV$ at 300K), low conductivity effective mass ($m_e^* = 0.014 m_0$ at the conduction band minimum at 300K)[26] and high refractive index ($n = 4$). Through moderate doping ($10^{16}$-$10^{19}$), the infrared permittivity of InSb can be continuously engineered between high refractive index ($\varepsilon_R$ ~ 16) and plasmonic ($\varepsilon_R<0$) regimes according to simple Drude models[27].

$$\epsilon = \epsilon' + i\epsilon'' = \epsilon_\infty \left(1 - \frac{\omega_p^2}{\omega^2 + i\Gamma\omega}\right); \quad \omega_p^2 = \frac{N_{free}e^2}{\epsilon_o \epsilon_\infty m_e^*}. \quad (1)$$

Exploiting this tremendous control over the IR optical properties, researchers can design "mesoscale" metamaterials where dielectric and metallic properties are seamlessly integrated within a single material system via MBE growth techniques[28,29]. As an example of *dynamic* control of these IR optical properties, consider the experimental reflectivity curves from an intrinsic (i.e. undoped) InSb wafer shown in Figure 1A. At room temperature (purple) the reflectivity is nearly constant with wavelength as expected for a simple dielectric interface with refractive index n~4. The drop in reflectivity at long wavelengths is consistent with a thermally generated free carrier density of $n_i \cong 4 \times 10^{17} cm^{-3}$. As the temperature is increased, the magnitude of the long wavelength reflectivity roll-off increases as well. By fitting these reflectivity curves (See Supplementary Information) we extract the electron concentration (Fig 1B) at each temperature. Note that the fits incorporate established models for the temperature-dependent band minimum electron effective mass (equation 2, where α = 3.9eV$^{-1}$ is the band non-parabolicity from the Kane model[30–32]). The extracted electron densities (red squares) closely match theoretical[33] predictions (equation 3):

$$m_e^*[T] = m_0^* \sqrt{1 + 4\alpha kT} \quad (2)$$

$$n_i[T] = 2.9 \times 10^{11}(2400 - T)^{0.75}(1 + 2.7 \times 10^{-4})T^{1.5} \exp\left(-\frac{0.129 - 1.5 \times 10^{-4}T}{kT}\right). \quad (3)$$

At long wavelengths, a 230K change in temperature can effect a larger than unity change in the refractive index (Fig 2D) with a nearly negligible impact on losses ($\Delta k < 2 \times 10^{-3}$). Note that the origin of this effect is fundamentally different than traditional thermo-optic effects in semiconductors; consequently the magnitude [$2n\frac{\partial n}{\partial T} \cong -0.2$ at λ = 13.5μm] is more than an order of magnitude larger than thermo-optic shifts reported for any group IV or III-V semiconductors[34] [Supplementary Information S2]. This large reduction in refractive index with temperature is in marked contrast to the behavior of doped InSb, described below.

The ability to engineer low loss ENZ materials by doping III-V semiconductors has enabled various applications[35–37]. However, the possibility of dynamically tuning ENZ wavelengths over large scales has not yet been demonstrated. ENZ substrates consisting of 1.2μm thick heavily n-doped (Te dopant) single crystal InSb films were MBE grown on lattice matched i-GaSb substrates (methods). Temperature dependent reflectivity curves are shown in Figure 2A. The reflectivity curves exhibit reflection minima at the onset of the plasma edge[38] ($\varepsilon_R$ ~ 1) followed by a longer wavelength bulk plasmon[39] at the epsilon-near-zero (ENZ) wavelength[40]. As the sample is heated, both features shift to longer wavelengths, indicating a reduction in the Drude weight and a corresponding increase in refractive index. The ENZ wavelength shifts by up to ~ 20% (Fig. 2B, red symbols), providing a new approach for generating tunable ENZ phenomena. Fitting reflectivity curves with a transfer-matrix method incorporating Drude models (Supplementary Information S1), we extract the scattering rate (Γ) and plasma frequency $\omega_p$ at each temperature. Unlike intrinsic InSb, where $\omega_p$ *increases* with temperature, the red-shifting features in Figure 2A are indicative of a *decreasing* plasma frequency.

To understand the origins of this decrease in plasma frequency we perform temperature dependent Hall measurements to independently measure the electron density (blue curve, Fig. 2B) and subsequently infer the elecron effective mass. Unlike the order of magnitude changes in carrier density for i-InSb, the carrier density for the doped films is relatively constant with temperature. The red-shift in plasma wavelength can only be explained by an increase in the electron effective mass. In heavily n-doped InSb, the Fermi level lies well above the conduction band minimum and the non-parabolic band strucutre plays a significant role. For instance, the extracted low-temperature (T=138K) effective mass (0.05) is more than three times larger than the value at the conduction band minimum (0.014$m_0$ at 300K). As the temperature is increased to 573K the effective mass increases by approximately 80% (Blue circles, Fig. 2C). The inferred effective mass shows excellent agreement with theoretical predictions[27,41] (blue line, equation 4) that account for the temperature dependent band gap (eqtn 5 for T< 600K),[42–44] band curvature (eqtn 2), and Fermi distribution:

$$m_n^*[T] = m_e^*[T]\sqrt{1 + \frac{1}{2}\left(\frac{3}{\pi}\right)^{\frac{2}{3}} \frac{h^2\, n[T]^{\frac{2}{3}}}{eE_g[T]\, m_e^*[T]}} \quad (4)$$

$$E_g[T] = 10^{-3}\left(235 - \frac{0.27 T^2}{T+106}\right) \quad (5)$$

The wavelength dependent change in the complex refractive index upon heating (138K to 573K) the n-doped InSb films is plotted in figure 2D. The real part of the refractive index increases by a maximum of 1.5 while the imaginary part of the refractive index decreases by 1.75 due to the increasing electron mass and accompanying reduction of the Drude response. The imaginary part of the dielectric constant varies between 0.5-0.7 at the ENZ wavelength across the whole temperature range, making it comparable with other low-loss ENZ materials.[45,46] Within a single material system we thus observe two different plasmonic tuning mechanisms: changing the free carrier concentration (intrinsic) or electron effective mass (doped). Below, we show that these distinct mechansims can alternately be used to achieve large resonance red-shifts or blue-shifts within a single resonator geometry.

The resonance wavelength of nano-antennas can be tuned by changing the refractive index of the resonator or of a supporting substrate.[47,48] Here, we show that i-InSb wire resonators fabricated on top of doped InSb layers demonstrate both classes of tunability, corresponding to the two distinct plasmon tuning mechanisms demonstrated above. Specifically, individual resonators exhibit 1) Transverse Electric (TE) resonances that are heavily coupled to an ENZ substrate and 2) Transverse Magnetic (TM) high-index Mie resonances that are almost independent of the underlying substrate. The distinct nature of these two resonances provides distinct temperature-dependent tuning capabilities.

**Geometric Dispersion of Dielectric Antennas on ENZ substrate**

Tunable resonators comprise low-loss dielectric (i-InSb, Fig 3A) wires (width w of 500nm to 5μm, 500μm long and 1μm tall) fabricated on heavily n-doped InSb films (methods). TE and TM polarized resonances of individual structures (Fig 3A) are measured with an FTIR microscope. The reflected signal from the resonator is normalized to an equivalent area background (substrate of same area of 50μm X 500μm and polarization)[18]. Experimentally measured dips in normalized reflection spectra (circles) show excellent agreement with scattering peaks in FDTD simulations (lines) for resonators of varying widths (Fig. 3C). TM (red) and TE (blue) polarized resonances show markedly different geometric dispersions (Supplementary Information S3). For the smallest width structures TE and TM resonances occur at similar wavelengths, where the substrate index is approaching zero. As width (w) increases (for a fixed height), TM modes show a strong dispersion of wavelength whereas TE modes exhibit far weaker size dispersion. For high-index cylindrical Mie resonators, the ratio of resonance wavelengths ($\frac{\lambda_1}{\lambda_2}$) is exactly equal to the square root of the ratio of cross-sectional area ($\sqrt{\frac{A_1}{A_2}} = \sqrt{\frac{w_1}{w_2}}$). A similar trend is expected for variations near square cross-section (w =1.2µm), and is consistent with the TM mode (8% increase in $\sqrt{w}$ ➔ 6% change in $\lambda_{res}$) but totally inconsistent with the TE mode (100% increase in $\sqrt{w}$ ➔ 13.2%- change in $\lambda_{res}$).

These differences in the geometric dispersion can be understood through an examination of the accompanying resonant field intensity profiles (Figure 3B). The TM resonant electric fields are almost completely localized within the resonator. The overlap of the electric field with the substrate is minimal and the geometric dispersion is almost independent of the substrate refractive index. On the other hand, the TE resonant electric fields are concentrated outside the resonator and are strongly coupled to the ENZ substrate. Similar to plasmonic resonators on ENZ substrates [ref], the mode thus exhibits a highly dispersive effective index $n_{eff}(\lambda)$. The resonance wavelegnth is proportional to the product of the width and effective index ($\lambda_{res} \propto n_{eff} w$); increasing width is offset by a decreasing $n_{eff}$ and the geometric dispersion is small with resonances getting "pinned" near the ENZ wavelength[39,49]. The difference between these two modes becomes particularly evident when we examine the temperature-dependent wavelength shifts.

**Thermal Dispersion of Dielectric Antennas on ENZ substrate**

Examples of temperature dependent reflection spectra for 2μm wide resonators on ENZ substrates are shown in Figure 4A. The TE resonance (solid lines) red-shifts with heating while the TM resonance blue-shifts. The TE resonance shifts are dominated by substrate refractive index changes; the increasing effective mass in the substrate (decreasing Drude weight) red-shifts the ENZ wavelength. The TE modes are pinned near this wavelength and red-shift in turn. The TE mode shifts agree well with simulations, exhibiting roughly linear resonance shifts across a large temperature range (80K to 573K) independent of resonator size (Fig. 4B). This dynamic tuning of the resonant wavelength is broadband (11-14μm) around the ENZ wavelength tuning range. Even though the TE modes are broad ($Q = \frac{\lambda_R}{\delta\lambda_{FWHM}}$ ~ 2.5-6) resonances, the large-magnitude index shift enables tuning by more than the line-width, an essential requirement of fully reconfigurable metadevices.

The TM resonances, on the other hand, exhibit more complex behavior across the full temperature range of these studies. The mode initially red-shifts at low temperatures, then demonstrates very rapid blue-shifts at temperatures above ~400 K (Fig. 4C). At low temperatures, the change in i-InSb free carrier concentration with heating is minimal. The small, observed red-shifts are mainly due to changes in the finite, negative permittivity of the doped substrate. At higher temperatures (T>400K), thermally generated free-carrier concentrations become significant (> $10^{17}$ cm$^{-3}$) and the resonator refractive index decreases quickly. Measurements of i-InSb on other substrates are expected to display larger shifts due to a lack of competition between substrate and resonator effects. Regardless, both effects are well accounted for in temperature-dependent simulations (dashed line), which show good agreement with experimental results.

Within a single material system, large refractive index shifts can be generated by either tuning of the electron effective mass (substrate tuning, visible in TE modes) or density (resonator tuning, visible in TM modes). These effects lead to a diversity of behaviors within even a single resonator geometry and are well characterized by simulations that incorporate temperature-dependent Drude models. The observed large-magnitude ($\Delta n>1$) refractive index shifts are far larger than conventional thermo-optic effects and suggest new avenues for highly tunable and reconfigurable mid-infrared Mie resonators, ENZ materials, and metasurfaces.

**Methods**

**Growth and Fabrication:** Molecular Beam Epitaxy growth of the layer structure (GaSb substrate/1.2μm Te-doped InSb/1μm i-InSb) was performed using a modified VG-V80H MBE system with a base pressure <5x10$^{-11}$ Torr. The undoped GaSb substrate was thermally desorbed under an Sb overpressure, after which a thin buffer of GaSb was grown. The substrate temperature was lowered to less than 380°C for the growth of Tellurium doped InSb, which was followed by growth of undoped InSb at the same growth temperature. After MBE growth, a 300nm thick film of SiO$_2$ was deposited using Plasma enhanced CVD to form a hard mask for subsequent dry etching. Photolithography was performed using a projection stepper aligner with SPR 90 as the photoresist (PR) along with Contrast Enhancing Mixture (CEM) top layer to increase verticality of the developed PR. The hard mask was etched using an inductively coupled plasma (ICP) dry etch using CHF$_3$/Ar gas mixture followed by a 10 minute Oxygen plasma clean at 250°C to remove any organic residue from the sample surface. The InSb was etched using a Reactive Ion Etching (RIE) process using a methane plasma and timed to stop at an etch depth of 1μm. The excess hard mask was removed using a wet etch with a buffered HF solution for 60s.

**Hall Measurement:** The temperature dependent hall measurements were performed on the MBE grown structure containing doped and un-doped layers of InSb from 2K to 400K using a Quantum Design Physical Property Measurement System in a Van-der-Paw geometry using a standard AC lock-in technique.

**FDTD Simulation:** The scattering cross section simulations for the single resonators were performed in a commercially available FDTD software[50] using the inbuilt cross section monitor and total-field-scattered field source. The symmetric and asymmetric boundary conditions were employed for TM and TE polarization simulations to accelerate the FDTD convergence. The two-dimensional simulations of wire cross section showed exceptional matching with the measured 3D wire resonator and the meshes around the resonator and ENZ substrate were minimized (1nmX1nm) to field divergence due to the vanishing dielectric constant. The near-field monitor was used to map the electric field intensity at the resonant wavelengths.

## Acknowledgement

We would like to thank Prof. Jim Allen for his useful insights during the course of this work. This work was supported by the Air Force Office of Scientific Research (FA9550-16-1-0393). We also acknowledge support from the Centre for Scientific Computing from the CNSI and NSF CNS-0960316. The authors would like to acknowledge the support of the UCSB MRL Shared Experimental Facilities which are supported by the MRSEC Program of the NSF under Award No. DMR 1121053; a member of the NSF-funded Materials Research Facilities Network. M.P. and C.J.P. would like to acknowledge support from the Vannevar Bush Faculty Fellowship program sponsored by the Basic Research Office of the Assistant Secretary of Defense for Research and Engineering and funded by the Office of Naval Research through grant N00014-15-1-2845. A part of this work was performed in the UCSB Nanofabrication Facility which is a part of the NSF funded National Nanotechnology Infrastructure Network.

## Author Contributions

P.P.I and J.A.S proposed and conceived the idea. M.P grew the InSb film and performed the Hall measurements under the supervision of C.J.P. P.P.I fabricated and measured the resonators and performed the electromagnetic simulations. P.P.I and J.A.S analyzed the data and all the authors contributed towards writing the manuscript.

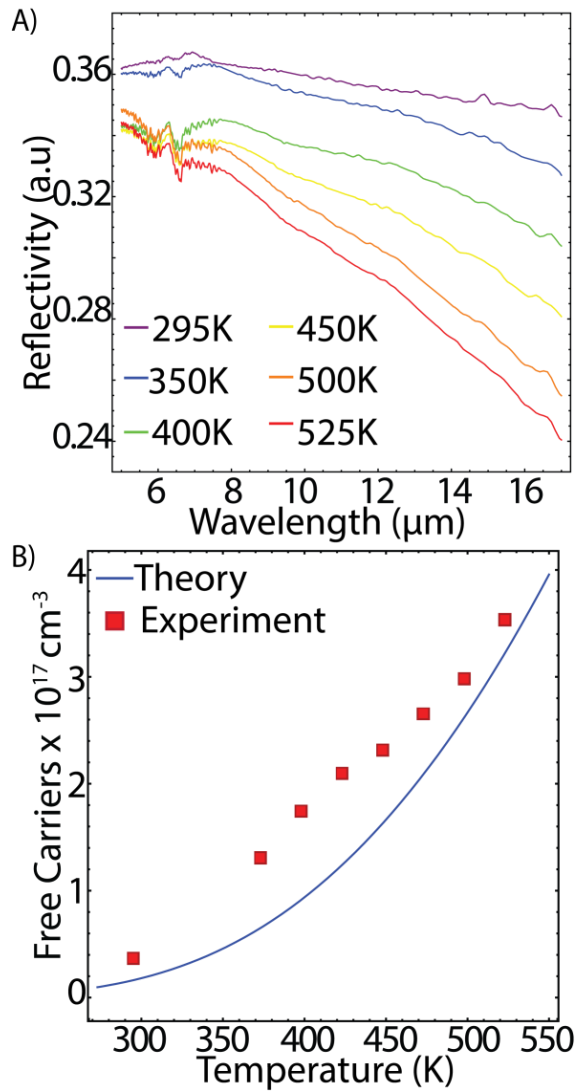

Figure 1. **A)** Reflection spectra of an intrinsic InSb film at various temperatures. The long-wavelength roll-off is due to Drude dispersion and increases with temperature. **B)** Temperature dependent free carrier concentrations inferred from fits to the curves in Fig 1A (square red dots) show good agreement with fit-free models of thermally activated carriers (blue).

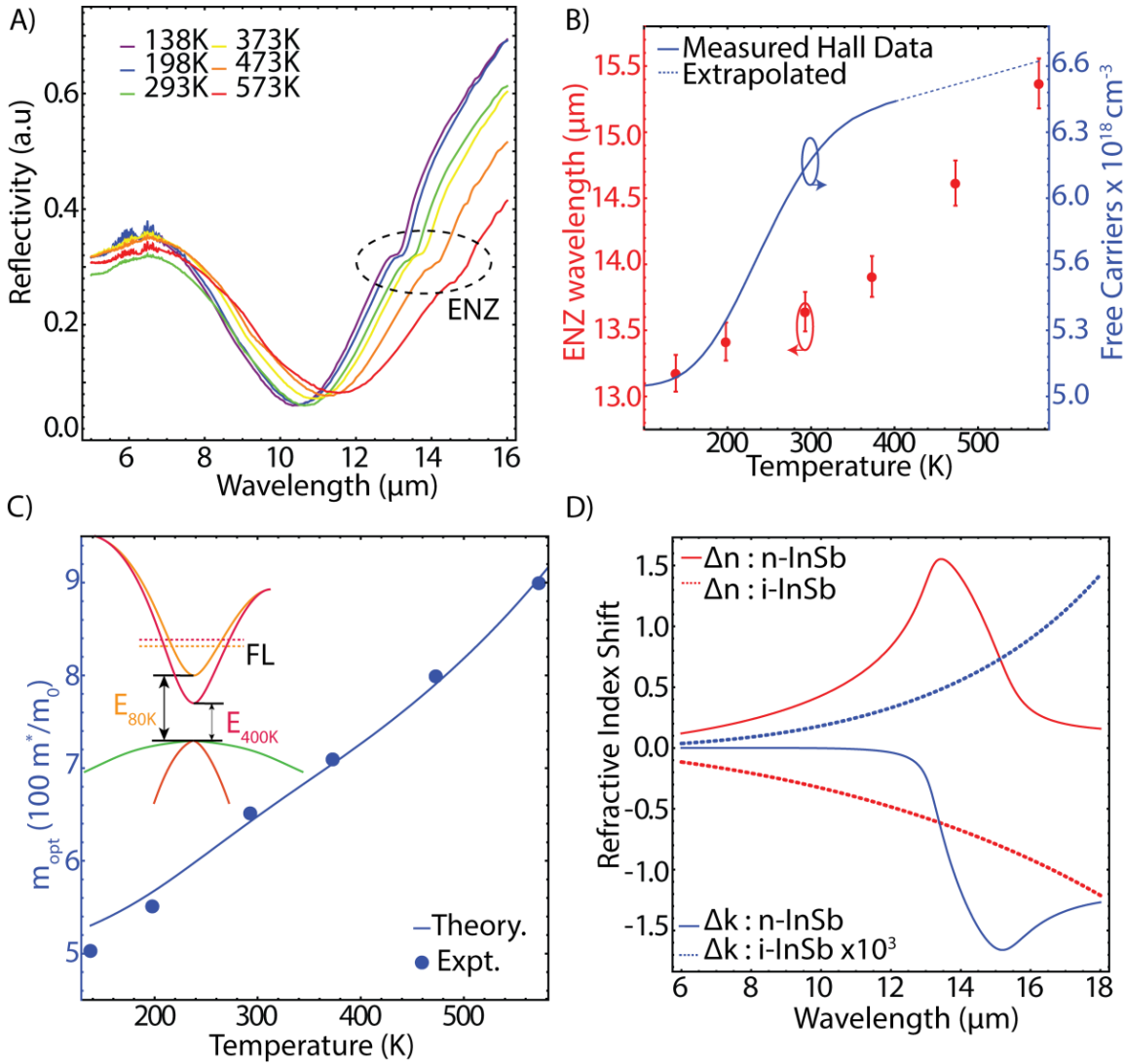

**A)** Reflection spectra of a highly doped InSb film at various temperatures. The dashed ellipse highlights a feature on the curve that marks the epsilon-near-zero (ENZ) wavelength. The ENZ wavelength red shifts with increasing temperature. **B)** Plot showing the thermal dispersion of $\lambda_{ENZ}$(red dots, left axis) and free carrier concentration (blue curve, right axis). $\lambda_{ENZ}$ is determined from transfer matrix modeling of plots in fig 2A. Free carrier concentration is determined from Hall measurements. **C)** Plot showing the experimentally determined optical effective mass ($m_{optical}$, blue dots) as a function of temperature. The effective mass increases by ~ 80%. The blue curve (left axis) indicates the predicted thermal dispersion of the electron effective mass for n-InSb according to eqns. 4 and 5. (inset) As temperature increases, changing band curvature and a shrinking bandgap leads to effective mass variation at the Fermi level (dashed lines) for highly doped InSb **D)** Plot showing the fitted refractive index change for both doped InSb (real part: red solid line; imaginary part: blue solid line) between 573K and 138K and intrinsic InSb (real part: red dashed line; imaginary part: blue dashed line) from 295K to 525K. Note that the real part

(imaginary part) increases (decreases) when doped InSb is heated, while the opposite is true for intrinsic InSb.

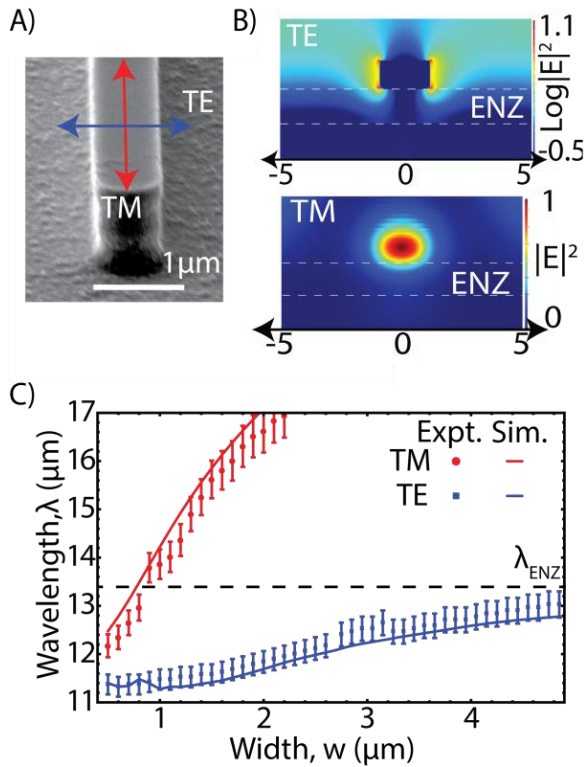

Figure 3: **A)** Scanning Electron Microscopy (SEM) images of the etched intrinsic InSb resonators of height 1µm, length 500µm and widths 700nm on the doped substrates. **B)** Plot showing the resonant electric field intensity profiles under TE and TM polarized illumination. The electric field intensity is concentrated outside the resonator, at the substrate-resonator edge for the TE polarized mode while the resonant fields are almost completely localized inside the resonator for the TM polarized mode. **C)** Plot showing the experimentally measured geometric dispersion of the TE (Blue squares) and TM (Red dots) resonances as function of resonator width. The solid lines (TE: blue & TM red) show the FDTD simulated peaks in scattering cross section of single resonators. The TE resonance exhibits a far weaker geometric dispersion and doesn't cross the ENZ wavelength of the substrate (black dashed line), indicating a highly substrate-coupled mode.

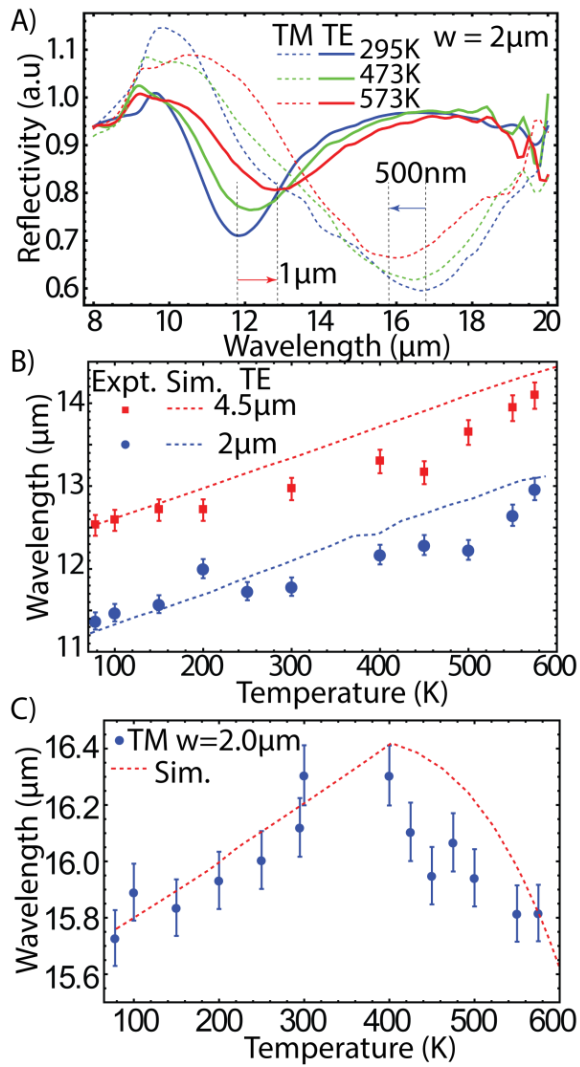

Figure 4. **A)** TE (solid) and TM (dashed) reflection spectra of a 2 μm wide resonator at different temperatures (295K: blue, 473K: green and 573K: red). The TM resonance red shifts while the TE resonance blue shifts with increasing temperature. **B)** Thermal dispersion of the experimentally measured (dots) and FDTD simulated (dashed lines) TE resonance as a function of temperature for 2μm (blue) and 4.5 μm (red) wide resonators. The resonances shift linearly by 1.5 μm from 80K to 573K for both resonators, in close agreement with FDTD simulations. **C)** Thermal dispersion of the TM resonance for a 2μm wide resonator. The resonance initially red shifts slowly due to substrate effects. Beyond 400K free carrier concentration increases rapidly leading to a reduced refractive index of the resonator and accompanying resonance blue-shifts. Both regimes of tunability are well accounted for in simulations (red dashed line).

1. Arbabi, A., Horie, Y., Bagheri, M. & Faraon, A. Dielectric metasurfaces for complete control of phase and polarization with subwavelength spatial resolution and high transmission. *Nat. Nanotechnol.* **10,** 937–943 (2015).

2. Yang, Y. *et al.* Dielectric Meta-Reflectarray for Broadband Linear Polarization Conversion and Optical Vortex Generation. *Nano Lett.* **14,** 1394–1399 (2014).

3. Meinzer, N., Barnes, W. L. & Hooper, I. R. Plasmonic meta-atoms and metasurfaces. *Nat. Photonics* **8,** 889–898 (2014).

4. Butakov, N. A. & Schuller, J. A. Designing Multipolar Resonances in Dielectric Metamaterials. *Sci. Rep.* **6,** (2016).

5. Yu, Y. F. *et al.* High-transmission dielectric metasurface with 2π phase control at visible wavelengths. *Laser Photonics Rev.* **9,** 412–418 (2015).

6. Khorasaninejad, M. *et al.* Achromatic Metasurface Lens at Telecommunication Wavelengths. *Nano Lett.* **15,** 5358–5362 (2015).

7. Lee, J. *et al.* Giant nonlinear response from plasmonic metasurfaces coupled to intersubband transitions. *Nature* **511,** 65–69 (2014).

8. Shcherbakov, M. R. *et al.* Ultrafast All-Optical Switching with Magnetic

Resonances in Nonlinear Dielectric Nanostructures. *Nano Lett.* **15,** 6985–6990 (2015).

9. Zheludev, N. I. & Plum, E. Reconfigurable nanomechanical photonic metamaterials. *Nat. Nanotechnol.* **11,** 16–22 (2016).

10. Zheludev, N. I. & Kivshar, Y. S. From metamaterials to metadevices. *Nat. Mater.* **11,** 917–924 (2012).

11. Michel, A.-K. U. *et al.* Reversible Optical Switching of Infrared Antenna Resonances with Ultrathin Phase-Change Layers Using Femtosecond Laser Pulses. *ACS Photonics* **1,** 833–839 (2014).

12. Iyer, P. P., Pendharkar, M. & Schuller, J. A. Electrically Reconfigurable Metasurfaces Using Heterojunction Resonators. *Adv. Opt. Mater.* **4,** 1582–1588 (2016).

13. Huang, Y.-W. *et al.* Gate-Tunable Conducting Oxide Metasurfaces. *Nano Lett.* **16,** 5319–5325 (2016).

14. Lu, F., Liu, B. & Shen, S. Infrared Wavefront Control Based on Graphene Metasurfaces. *Adv. Opt. Mater.* **2,** 794–799 (2014).

15. Yao, Y. *et al.* Electrically tunable metasurface perfect absorbers for ultra-thin

mid-infrared optical modulators. *Nano Lett.* **14,** 6526–6532 (2014).

16. Wang, Q. *et al.* Optically reconfigurable metasurfaces and photonic devices based on phase change materials. *Nat. Photonics* **10,** 60–65 (2015).

17. Li, Z. *et al.* Correlated Perovskites as a New Platform for Super-Broadband-Tunable Photonics. *Adv. Mater.* **28,** 9117–9125 (2016).

18. Lewi, T., Iyer, P. P., Butakov, N. A., Mikhailovsky, A. A. & Schuller, J. A. Widely Tunable Infrared Antennas Using Free Carrier Refraction. *Nano Lett.* **15,** 8188–93 (2015).

19. Liu, X. & Padilla, W. J. Thermochromic Infrared Metamaterials. *Adv. Mater.* **28,** 871–875 (2016).

20. Kuznetsov, A. I., Miroshnichenko, A. E., Fu, Y. H., Zhang, J. & Luk'yanchuk, B. Magnetic light. *Scientific Reports* **2,** (2012).

21. Schuller, J. A. & Brongersma, M. L. General properties of dielectric optical antennas. *Opt. Express* **17,** 24084–95 (2009).

22. Iyer, P. P., Butakov, N. A. & Schuller, J. A. Reconfigurable Semiconductor Phased-Array Metasurfaces. *ACS Photonics* **2,** 1077–1084 (2015).

23. Baranov, D. G., Makarov, S. V., Krasnok, A. E., Belov, P. A. & Alù, A. Tuning of

near- and far-field properties of all-dielectric dimer nanoantennas via ultrafast electron-hole plasma photoexcitation. *Laser Photon. Rev.* **10,** 1009–1015 (2016).

24. Guo, P., Schaller, R. D., Ketterson, J. B. & Chang, R. P. H. Ultrafast switching of tunable infrared plasmons in indium tin oxide nanorod arrays with large absolute amplitude. *Nat. Photonics* **10,** 267–273 (2016).

25. Wolf, O., Campione, S., Kim, J. & Brener, I. Spectral filtering using active metasurfaces compatible with narrow bandgap III-V infrared detectors. *Opt. Express* **24,** 21512 (2016).

26. Zawadzki, W. Electron transport phenomena in small-gap semiconductors. *Adv. Phys.* **23,** 435–522 (1974).

27. Law, S., Liu, R. & Wasserman, D. Doped semiconductors with band-edge plasma frequencies. *J. Vac. Sci. Technol. B, Nanotechnol. Microelectron. Mater. Process. Meas. Phenom.* **32,** 52601 (2014).

28. Wei, D. *et al.* Single-material semiconductor hyperbolic metamaterials. *Opt. Express* **24,** 8735 (2016).

29. Law, S. *et al.* All-Semiconductor Negative-Index Plasmonic Absorbers. *Phys. Rev. Lett.* **112,** 17401 (2014).

30. van Welzenis, R. G. G. & Ridley, B. K. K. On the properties of InSb quantum wells. *Solid. State. Electron.* **27,** 113–120 (1984).

31. Carelli, G. *et al.* Temperature Dependence of InSb Reflectivity in the Far Infrared: Determination of the Electron Effective Mass. *Int. J. Infrared Millimeter Waves* **19,** 1191–1199 (1998).

32. Liu, P. Y. & Maan, J. C. Optical properties of InSb between 300 and 700 K. II. Magneto-optical experiments. *Phys. Rev. B* **47,** 16279–16285 (1993).

33. Oszwałldowski, M. & Zimpel, M. Temperature dependence of intrinsic carrier concentration and density of states effective mass of heavy holes in InSb. *J. Phys. Chem. Solids* **49,** 1179–1185 (1988).

34. Ghosh, G. *Handbook of Thermo-Optic Coefficients of Optical Materials with Applications*. *Academic Press* (Academic Press, 1998).

35. Alù, A., Silveirinha, M. G., Salandrino, A. & Engheta, N. Epsilon-near-zero metamaterials and electromagnetic sources: Tailoring the radiation phase pattern. *Phys. Rev. B* **75,** 155410 (2007).

36. Dyachenko, P. N. *et al.* Controlling thermal emission with refractory epsilon-near-zero metamaterials via topological transitions. *Nat. Commun.* **7,** 11809 (2016).

37. Liberal, I. & Engheta, N. Nonradiating and radiating modes excited by quantum emitters in open epsilon-near-zero cavities. *Sci. Adv.* **2,** (2016).

38. Spitzer, W. G. & Fan, H. Y. Determination of Optical Constants and Carrier Effective Mass of Semiconductors. *Phys. Rev.* **106,** 882–890 (1957).

39. Schulz, S. A. *et al.* Optical response of dipole antennas on an epsilon-near-zero substrate. *Phys. Rev. A* **93,** 63846 (2016).

40. James Allen, S., Raghavan, S., Schumann, T., Law, K.-M. & Stemmer, S. Conduction band edge effective mass of La-doped BaSnO3. *Appl. Phys. Lett.* **108,** 252107 (2016).

41. Kolodziejczak, J., Byszewski, P., Kołodziejczak, J. & Zukotyński, S. The Thermoelectric Power in InSb in the Presence of an External Magnetic Field. *Phys. status solidi* **1880,** 1880–1884 (1963).

42. Vurgaftman, I., Meyer, J. R. & Ram-Mohan, L. R. Band parameters for III-V compound semiconductors and their alloys. *J. Appl. Phys.* **89,** 5815–5875 (2001).

43. Littler, C. L. & Seiler, D. G. Temperature dependence of the energy gap of InSb using nonlinear optical techniques. *Appl. Phys. Lett.* **46,** 986 (1985).

44. Liu, P. Y. & Maan, J. C. Optical properties of InSb between 300 and 700 K. I. Temperature dependence of the energy gap. *Phys. Rev. B* **47,** 16274–16278 (1993).

45. Javani, M. H. & Stockman, M. I. Real and Imaginary Properties of Epsilon-Near-Zero Materials. *Phys. Rev. Lett.* **107404,** 1–6 (2016).

46. Naik, G. V., Shalaev, V. M. & Boltasseva, A. Alternative plasmonic materials: Beyond gold and silver. *Adv. Mater.* **25,** 3264–3294 (2013).

47. Jun, Y. C. *et al.* Epsilon-Near-Zero Strong Coupling in Metamaterial-Semiconductor Hybrid Structures. *Nano Lett.* **13,** 5391–5396 (2013).

48. Sautter, J. *et al.* Active Tuning of All-Dielectric Metasurfaces. *ACS Nano* **9,** 4308–4315 (2015).

49. Kim, J. *et al.* Role of epsilon-near-zero substrates in the optical response of plasmonic antennas. *Optica* **3,** 4–6 (2016).

50. FDTD Solutions Lumerical.com. Lumerical Solutions Inc. (2015).

# Supplementary Information

## Ultra-wide plasmonic tuning of semiconductor metasurface resonators on epsilon near zero media

Prasad.P.Iyer[1], Mihir Pendharkar[1], Chris J. Palmstrøm[1,2], Jon A. Schuller[1]

1. Electrical and Computer Engineering Department, University of California Santa Barbara, CA
2. Material Science and Engineering Department, University of California Santa Barbara, CA

1. **Transfer Matrix Model Fitting for the Drude parameters**

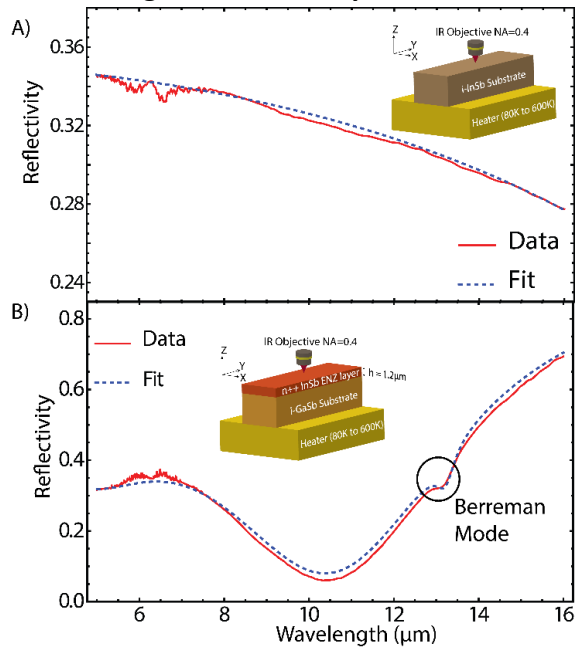

Figure S1: **A)** Reflectivity spectrum from single crystal substrate of intrinsic InSb (Red, solid) and the analytical Fresnel reflection fit curve for an infinite InSb slab (Blue, Dashed). The reflectivity rolls off at long wavelengths due to Drude dispersion from thermally activated carriers **B)** Reflectivity spectrum from a thin film of doped InSb on GaSb (Red, solid) and the Transfer matrix fit curve for the same stack under off-normal incidence (Blue, Dashed). The small "kink" in the data around 13.5µm, highlighted by the black circle, shows the Berreman mode near the ENZ wavelength. The inset sketch shows the measurement setup.

The intrinsic InSb reflectivity curves (normalized to Au) are fit based on simple Fresnel equations for a semi-infinite slab of dielectric whose refractive index model includes free-carrier Drude dispersion. Reflectivity roll off at higher wavelength is caused by thermally-generated free-carriers throughout the whole substrate. From these fits we determine the electron density as a function of temperature. The Mid-infrared reflection spectra (normalized to Au) of doped InSb is fit to a transfer matrix model of a single dielectric film on a semi-infinite slab of GaSb. The reflection spectra exhibits distinct characteristics that enable unique fits of the Drude (equation 1) plasma frequency ($\omega_p$) and scattering rate ($\Gamma$) . The dip in the reflection at 11µm marks the plasma edge (n~1) while the "kink" on the rising edge of the reflection curve at 13µm marks a Ferrel-Berreman mode near the ENZ wavelength (n~0). The peak in the reflection curve 6µm enables us to determine the thickness of the thin film (1.2µm), assuming GaSb substrate index at 3.8 and $\epsilon_\infty$ of InSb at 15.68. As the plasma frequency increases the plasma edge dip blue shifts. Reflection measurements are performed in an Infrared microscope with 15X objective, ensuring off-normal incidence angles up to ~20°. This enables coupling to the Ferrel-Berreman bulk plasmon mode of the thin film. Fits of the scattering rate ($\Gamma$) are primarily sensitive to the slope of the rising edge and sharpness (or line-width) of the Ferrel-Berreman feature. As the scattering rate increases, the slope of the rising reflectivity curve for λ > plasma edge dip decreases along with the line-width of the kink. Thus the thermal dispersion of the plasma frequency and the scattering rate are determined with little correlation between the two parameters. Using measured free carrier concentration from temperature dependent Hall measurements, the electron effective mass can be determined uniquely.

2. **Thermo-optic Coefficients for Traditional Semiconductors in the LWIR wavelength**

| Semiconductor | $2n \frac{\partial n}{\partial T} \times 10^{-4}/K$ |
|---|---|
| Si | 10 |
| Ge | 35 |
| InAs | 22 |
| GaAs | 18 |
| PbSe | -90 |
| PbTe | -145 |
| i-InSb* | -2080 |

*Intrinsic InSb thermo-optic shift reported in this work is based on the thermal free carrier generation based refractive index shift at 13.5µm.

3. **Geometric Dispersion of the resonances from FDTD.**

The scattering cross-section from a single wire resonator is measured using power-monitor outside the total field scattered field (TFSF) plane wave polarized source. The dips in measured reflection curves match closely with the peaks in simulated cross section curves (Fig 3B). There is a slight offset for TE (red shift) and TM (blue shift) polarized resonances measured due to a substrate effect introduced through the normalized spectral measurements.

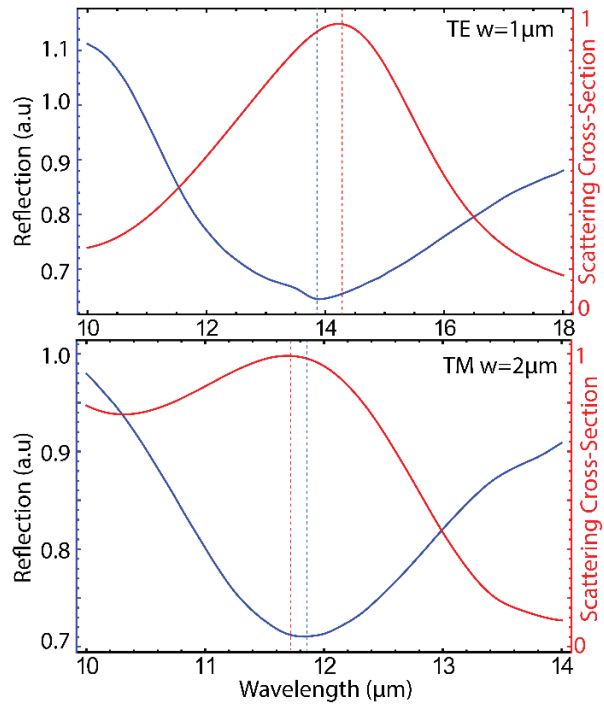

Figure S2: A) Experimentally measured reflectivity curve (Blue, left axis) and simulated normalized scattering cross-section curve (Red, right axis) under TE polarization for a resonator of width 1μm B) Experimentally measured reflectivity curve (Blue, left axis) and simulated normalized scattering cross-section curve (Red, right axis) under TM polarization for a resonator of width 2μm.